# Distributed Sensing of Single Mode Fibers with Correlation Techniques

Florian Azendorf
Advanced Technology
ADVA Optical Networking SE
Meiningen, Germany
fazendorf@adva.com

André Sandmann
Advanced Technology
ADVA Optical Networking SE
Meiningen, Germany
asandmann@adva.com

Michael Eiselt
Advanced Technology
ADVA Optical Networking SE
Meiningen, Germany
meiselt@adva.com

Bernhard Schmauss
Institute of Microwaves and Photonics
Friedrich-Alexander-Universität
Erlangen-Nürnberg
Erlangen, Germany
bernhard.schmauss@fau.de

**Abstract—** In this paper, we report on the development progress of correlation-based optical time domain reflectometry (OTDR). Substituting the direct detection receiver with a coherent receiver enables to extract the phase and polarization information of the reflected signal. Furthermore, due to the mixing of a weak probe signal with a strong local oscillator the sensitivity of the receiver improved. This improvement was demonstrated by analyzing the reflection from an angled physical contact (APC) connector. To further quantify the improvements, we compare the direct detection correlation OTDR (C-OTDR) with the coherent detection correlation OTDR (CC-OTDR) with respect to the spatial and amplitude resolution. Keywords—distributed sensing, OTDR, correlation techniques, phase noise,

## I. Introduction

The usage of optical fibers as a sensing medium has attracted considerable interest due to the promising properties of fibers like high sensitivity, operation in harsh environments and the immunity to electromagnetism. In recent years, also network carriers have gained interest to use their deployed fibers as combined transport and sensing media. The carrier obtains information about the surrounding fiber environment, which can be used for operational purposes or to gain further revenue. As another approach besides distributed sensing of the telecom fiber itself, fiber Bragg gratings (FBG) can be inscribed in the core for a quasi-distributed sensing. A promising technique is the use of ultra-weak FBGs (UWFBG) [1], which have a reflectivity in the order of -30 dB per grating and can thus be arranged in series to form a sensor array. The reflection of each grating leads to a discrete sampling point along the fiber, from which the reflection is significantly higher than Rayleigh backscattering. However, when thousands of gratings are cascaded, the probe signal experiences significantly higher attenuation, as compared to Rayleigh backscattering. An appropriate means to interrogate deployed fibers is OTDR. The technique measures the Rayleigh backscattering and Fresnel reflections by inserting an optical pulse into the fiber under test. Systems based on direct detection have been used for monitoring in long-haul fiber networks for more than 30 years [2]. These systems analyze the amplitude of the reflected and backscattered light. In addition to the amplitude, coherent detection enables the measurement of phase and polarization of the optical signal, being backscattered over the fiber length. Consequently, it is feasible to obtain further information about the fiber and the environment [3]. In both cases, a rectangular shaped pulse is used to probe the fiber. Alternatively, code sequences can be used as the probe signal, improving the spatial resolution [4], [5]. In this work we will compare, the correlation-based direct detection method with the coherent method. In section II, the measurement principle and setup of the coherent correlation technique are introduced. Section III, shows the result of the fingerprint measurement, and compares both methods regarding their spatial resolution and amplitude sensitivity. In section IV, we will discuss the laser phase noise as one of the main issues impacting the measurement, when using telecommunication components. Finally, we conclude our findings and present an outlook in section V.

## II. General Measurement Setup

The general measurement setup of the coherent correlation OTDR is illustrated in Figure 1. A tunable laser with a typical linewidth of 25 kHz generates a continuous wave signal. One part is used as the local oscillator (LO) of the coherent receiver, while the other part is fed into a Mach-Zehnder modulator (MZM). A bipolar sequence is modulated onto the carrier, and the probe signal is sent into the fiber under test (FuT) via an optical circulator. To reduce the influence of undesired strong reflections at the input and output of the FuT, angled physical connectors (APC) are used. The reflected and backscattered signals are received with a dual-polarization integrated coherent receiver and recorded with a real-time oscilloscope. In a single shot, multiple frames of the sequence are transmitted and consecutively recorded. Similar to [5], each frame consists either of a pseudo random binary sequence (PRBS) or of two Golay sequences, followed by a zero padding. The overall frame length corresponds to the round-trip time of the FuT to obtain unambiguous results. Four signals are recorded at the output of the coherent receiver, representing the in-phase and quadrature components for x- and y-polarizations.

A cross-correlation between the transmitted and the received sequences is then performed, improving the signal-to-noise ratio and the spatial resolution. Here, complementary Golay sequences exhibit perfect acyclic auto-correlation properties, if both sequences are used and their respective correlation functions are added. After cross-correlation in the testbed setup, the frames are further processed to extract amplitude, phase, and polarization of the backscattered signals.

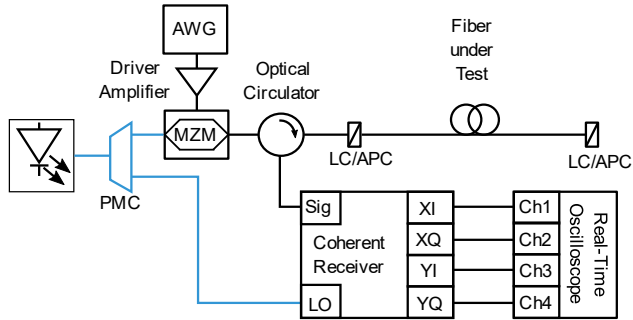

Figure 1 General schematic of the Coherent Correlation-OTDR. Blue color denotes polarization maintaining fibers.

### III. MEASUREMENTS AND COMPARISON

#### A. Fingerprint of a single mode optical fiber

Rayleigh backscattering is a result of the inhomogeneities along the fiber. These inhomogeneities originate from the amorphous character of glass. By probing the fiber with a pulse or sequence, the backscattering trace can be observed, which is a coherent combination of the backscattered fields from the individual inhomogeneities, and which is constant under steady conditions. This trace is unique for each fiber, and therefore it can be denoted as a fingerprint. The fingerprint of a fiber with length 200 m was measured with the CC-OTDR. In the experiment, we used a probe bit rate of 1.25 Gbit/s, a sampling rate of 25 GS/s, and a 511-bit PRBS. 500 consecutive frames were recorded. The recorded signals were filtered with a low pass filter and the cross correlation with the transmitted sequence was performed. To obtain the fingerprint, the absolute square of the individual correlation signals for x- and y-polarization are added. The results of the investigation are depicted in Figure 2 starting at fiber length 48.3 m.

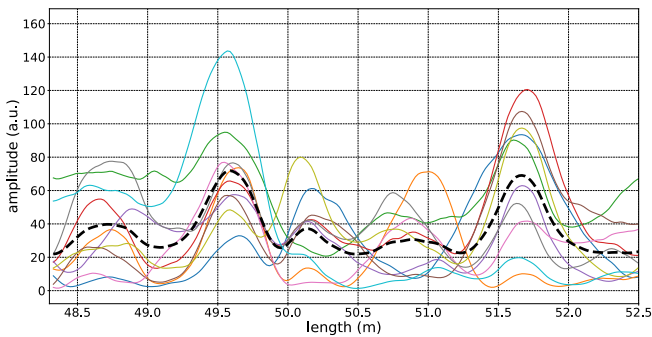

Figure 2 Fingerprint of individual frames (solid line) and averaged fingerprint of the ten frames (dashed line).

In the figure, the amplitude of each fingerprint is depicted as a function of fiber length for ten consecutive frames. Each solid line represents a recorded trace, while the dashed line shows the average trace of the ten frames. It can be seen that there is a good overlap of the peak locations of the ten frames with the average trace at 49.57 m, and 51.67 m. Additionally, weak peaks were observed at 48.715 m, 50.135 m, and 50.87 m. The variations of the amplitude are presumably caused by vibrations and noise in the laboratory.

#### B. Sensitivity

It has been highlighted in the literature that coherent detection improves the sensitivity compared to the typical OTDR with direct detection [6, 7]. The sensitivity describes the scale factor between measurand (i.e., the reflected and backscattered optical power) and the received electrical signal, and it defines, which scattering, or reflection events are resolvable. It should be noted that the term sensitivity can be misunderstood with measurement resolution. The latter is the ability of the measurement technique to distinguish small changes in the quantity that is measured [8]. The probe bit rate in the experiment was 5 Gbit/s, two 2048-bit Golay sequences were used, and the sampling rate was 25 GS/s. As fiber under test, three patch cords with different lengths were cascaded. The results of the experiment are illustrated in Figure 3 showing the averaged absolute square of the correlation signals, as a function of the fiber length.

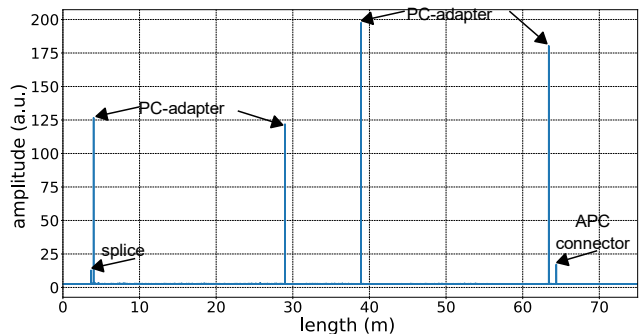

Figure 3 Averaged CC-OTDR fingerprint showing the three cascaded jumper fibers and several reflections from PC and APC connectors and a splice.

The figure shows several reflection peaks at the interface between the fibers with lengths of 25 m, 10 m, and 25 m. Furthermore, the results demonstrate the feasibility to measure APC connector reflections as well as reflections caused by splices, with both events having low reflectance in the order of -60 dB [2]. The experiment was repeated with the direct detection C-OTDR similar to our previous work [5]. The coherent receiver shown in Figure 1 was substituted by a photodiode/amplifier (PIN/TIA) combination. Since no LO is needed in the direct detection setup, the laser was directly connected to the MZM. Additionally, two Golay sequences were used to probe the fiber. The results of this experiment are illustrated in Figure 4. In the figure, the sum of the correlation functions is shown. It can be seen that the PIN/TIA combination of the direct detection method has a significantly lower sensitivity, which is noticeable since no events are visible. The reason is that the reflections of the adapters and connectors are below -46 dBm, which was measured at the optical circulator. Another reason is the vertical resolution of the real-time oscilloscope with 10 mV/div and its 8-bit quantization.

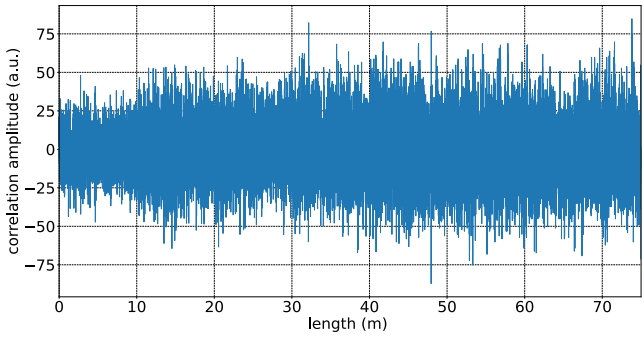

Figure 4 Amplitude of the correlation trace for direct detect C-OTDR and the same FuT as in Figure 3.

*C. Spatial Resolution*

Another important parameter of OTDR measurement techniques is the spatial resolution. This parameter describes the capability of an OTDR to distinguish between reflective events with close proximity. In general, the parameter depends on the probe pulse width and the bandwidth as well as the sampling rate of the receiver structure. By considering all parameters, the spatial resolution of both techniques can be estimated and measured. The spatial resolution of the C-OTDR with direct detection was determined by using two fibers with roughly the same length as FuT coupled together with a 3 dB coupler. In the experiment the probe bit rate was varied between 10 Gbit/s, 5 Gbit/s and 2.5 Gbit/s, and the sampling rate was 50 GS/s. The overall bandwidth of the photodiode and the real-time oscilloscope was 10 GHz. Consequently, the temporal resolution should be 100 ps, 200ps and 400 ps, corresponding to a spatial resolution of 10 mm, 20 mm, and 40 mm, respectively. The results in Figure 5 show the measured trace with the different probe bit rates.

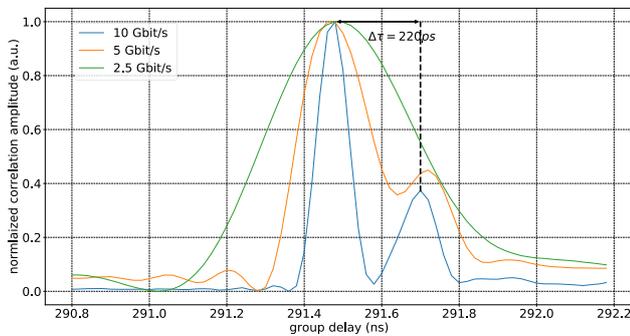

Figure 5 Spatial resolution analysis with different probe bit rates of the correlation sequence with direct detection.

From the measurements it can be seen that with 5 Gbit/s and 10 Gbit/s data rates it is feasible to measure a difference of 220 ps, which corresponds to a length of 2.2 cm between two reflection events. To investigate the resolution of the coherent correlation OTDR, a UWFBG array inscribed in a 106 m fiber was used with a grating separation of 50 mm. Each grating had a length of 10 mm, and the reflectivity was -30 dB each. The results of the experiment are illustrated in Figure 6 showing the correlation signal of the first 100 gratings with different amplitudes. The inset of the figure shows the whole fiber length with 2000 gratings, sampled at a frequency of 193.4 THz for that particular measurement. The results show the reflection for one sample frequency of each FBG. Since the Bragg wavelength is different for each FBG the correlation amplitude differs.

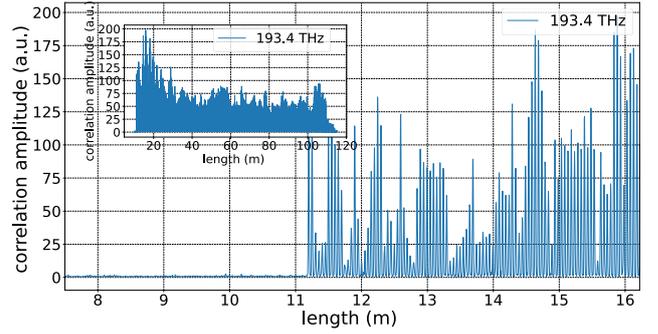

Figure 6 Correlation signal of the first 100 gratings of the FuT probed with a carrier frequency of 193.4 THz. Inset: Signal of the whole FuT with 2000 grating.

## IV. LASER PHASE NOISE INFLUENCE

In order to obtain optimum results from the correlation in the coherent detection setup, the laser phase difference between the LO and the reflected signal needs to be stable over the received sequence pattern. Hereinafter, an experiment is conducted that highlights the effect of laser phase noise on the correlation result. Two fibers under test are interrogated with the CC-OTDR setup. The first is a short fiber of 8 meter length with an open PC connector at the fiber end (i.e. transition from silica to air) to generate a strong reflection. Secondly, a 10.14 km fiber is used, which has the same open connector at the fiber end. A set of Golay sequences of length 4096 bits is transmitted with a symbol rate of 1.25 GBaud in a time-multiplexed fashion. The respective correlation functions of each complementary sequence are combined to obtain the total correlation trace. The resulting normalized total correlation magnitudes of the end-face reflections for the two fiber lengths are shown in Figure 7.

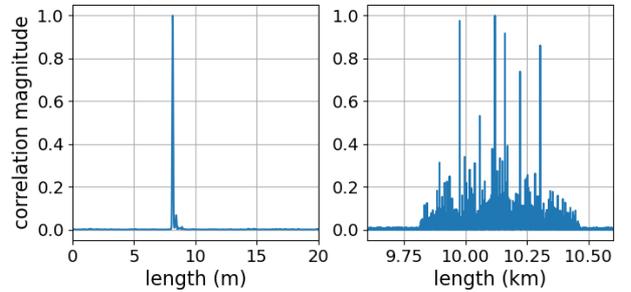

Figure 7 Normalized correlation magnitude (x-polarization only) of an open PC connector after 8 meters (left) and 10.14 km (right)

As expected for the short fiber of 8 m length, a clear correlation peak is visible. In contrast, the reflection originating from the end of the 10.14 km fiber results in a correlation pattern with multiple peaks and high side lobe noise. The latter is a result of a fast drift of the phase difference between the reflected signal and the LO over the received correlation sequence. Since a single-laser setup with self-homodyne detection is used, as shown in Figure 1, the superimposed LO and signal at the receiver are affected by different laser phase noise evolutions due to their different delays. In addition, the laser that is implemented in this experiment exhibits a slow modulation of its instantaneous frequency. Both effects combined lead to the fast drift of the phase difference when considering the reflection from 10.14 km. Considering a single code sequence, the phase difference needs to be stable over the received sequence. Using two complementary sequences, the phase difference needs to be stable such that both sequences are measured with the same phase difference.

## V. SUMMARY AND CONCLUSION

In this work, we presented a comparison between a direct detection and a coherent correlation OTDR method. It was shown that the sensitivity of the method was improved due to coherent detection. Furthermore, the spatial resolutions were compared, and it was shown that the parameter depends on the probe bit rate and the overall bandwidth of the receiver.

## ACKNOWLEDGMENT

This work was partly funded by the German Federal Ministry of Education and Research (FKZ16KIS1279K) in the framework of the CELTIC-NEXT project AI-NET-Protect (Project ID C2019/3-4).

## REFERENCES

[1] M. Wu, C. Li, X. Fan, C. Liao and Z. He, "Large-scaled multiplexed weak reflector array fabricated with a femtosceond laser for a fiberoptic quasi-distributed acoustic sensing system," *Optical Letters,* pp. 3685-3688, 1 Juli 2020.

[2] D. Anderson, L. Johnson and F. Bell, Troubleshooting Optical Fiber Networks-Understanding and Using Your Optical Time-Domain Reflectometer, Academic Press Elsevier, 2004.

[3] E. Ip, J. Fang, Y. Li, Q. Wang, M. F. Huang, M. Salemi and Y. K. Huang, "Distributed Fiber Sensor Network Using Telecom Cables as Sensing Media Technology Advancements and Applications," *Journal of Optical Communications and Networking ,* vol. 14, no. 1, pp. A61-A68, 2022.

[4] C. Dorize and E. Awwad, "Enhancing the Performance of Coherent OTDR Systems with Polarization Diversity Complementary Codes," *Optics Express,* vol. 26, pp. 12878-12890, 2018.

[5] F. Azendorf, A. Dochhan and M. H. Eiselt, "Accurate Single-Ended Measurement of Propagation Delay in Fiber Using Correlation Optical Time Domain Reflectometry," *Journal of Lightwave Technology,* pp. 5744-5752, 15 September 2021.

[6] A. H. Hartog, An Introduction to Distributed Optical Fibre Sensors, Boca Raton: CRC Press, 2017.

[7] L. S. Grattan and B. T. Meggitt, Optical Fiber Sensor Technology: Advanced Applications-Bragg Gratings and Distributed Sensors, Boston, MA: Springer, 200.

[8] JCGM, "International Vocabulary of Metrology–Basic and General Concepts and Associated Terms," in *Chemistry International*, 2012.